 \documentclass[11pt]{article}

\usepackage[round]{natbib}   

\usepackage[english]{babel}

\usepackage[utf8]{inputenc} 
\usepackage[T1]{fontenc}
\usepackage{lmodern}

\usepackage{pifont}     

\usepackage{graphicx}
\usepackage{amsmath}

\usepackage{enumitem}

\usepackage{float}
\usepackage{ifthen}
\usepackage{hyperref}

\usepackage{color}

\begin{document}

\begin{center}
{\bf \Large \color{blue} 
The third law of thermodynamics or an absolute definition
}
\\ 
\vspace*{2mm}
{\bf \Large \color{blue}  
for Entropy. Part 2 : definitions and applications
}
\\ 
\vspace*{2mm}
{\bf \Large \color{blue}  
in Meteorology and Climate.
}
\\ 
\vspace*{4mm}
{\bf \large \color{black}  
Accepted for the revue: ``\textbf{\emph{La Météorologie}}\,'' / Version 2 - \today.
}
\\ \vspace*{5mm}
{\Large by Pascal Marquet. M\'et\'eo-France, CNRM/GMAP, Toulouse.}
\end{center}


\begin{center}
{\Large \bf Abstract}
This paper is the second part of a previous paper (Marquet, 2019) dealing with the need to define the entropy with an absolute way, by using the third law of thermodynamics. In this second part it is shown that there is a need and interest to define a potential temperature which is a synonym of the moist-air absolute entropy, with several possible novel applications to study meteorology and climate processes.
\end{center}
\vspace*{-2mm}

\section{\underline{\Large Entropy and potential temperatures in Meteorology}}
\label{section_1}
\vspace{-1mm}

Without revealing too much of the end of the story, the purpose of this study is to show that we can define a ``potential temperature'' that is completely synonymous with the moist-air entropy, whatever the conditions of local temperature, pressure and composition of the atmosphere (vapour moisture and possible liquid or icy condensates) may be. 
We can therefore ask ourselves a first questions: what is the notion of ``potential temperature''\,? and what are the links that can exist with entropy, a notion that was invented by Clausius in 1865\,? (see Part 1).

For once, we can answer this egg-and-chicken problem\:: it is clearly the notion of ``potential temperature'' which preceded that of the entropy in meteorology.
In fact, the idea that quantities could be preserved during vertical motions in the atmosphere was established by Poisson as early as 1833, more than 30 years before the discovery of the entropy by Clausius.

By using modern notations, the consequence of the equations~(6) of Poisson (Proposition 638, p.647) is that the quantity $ T \, p^{\,(1- \gamma) /\gamma}$ is conserved for adiabatic transformations, where $\gamma=c_p/c_v$ a constant close to $1.4$ for atmospheric gases.
This leads to the adiabatic conservation for what is nowadays called the ``potential temperature''\:: $ \theta = T \, (p_0/p)^{\kappa}$, where $p_0=1000$~hPa is a constant pressure and $\kappa = (\gamma-1) /\gamma \approx 0.2857$.

\begin{figure}[hbt]
\centering
\includegraphics[width=0.99\linewidth,angle=0,clip=true]{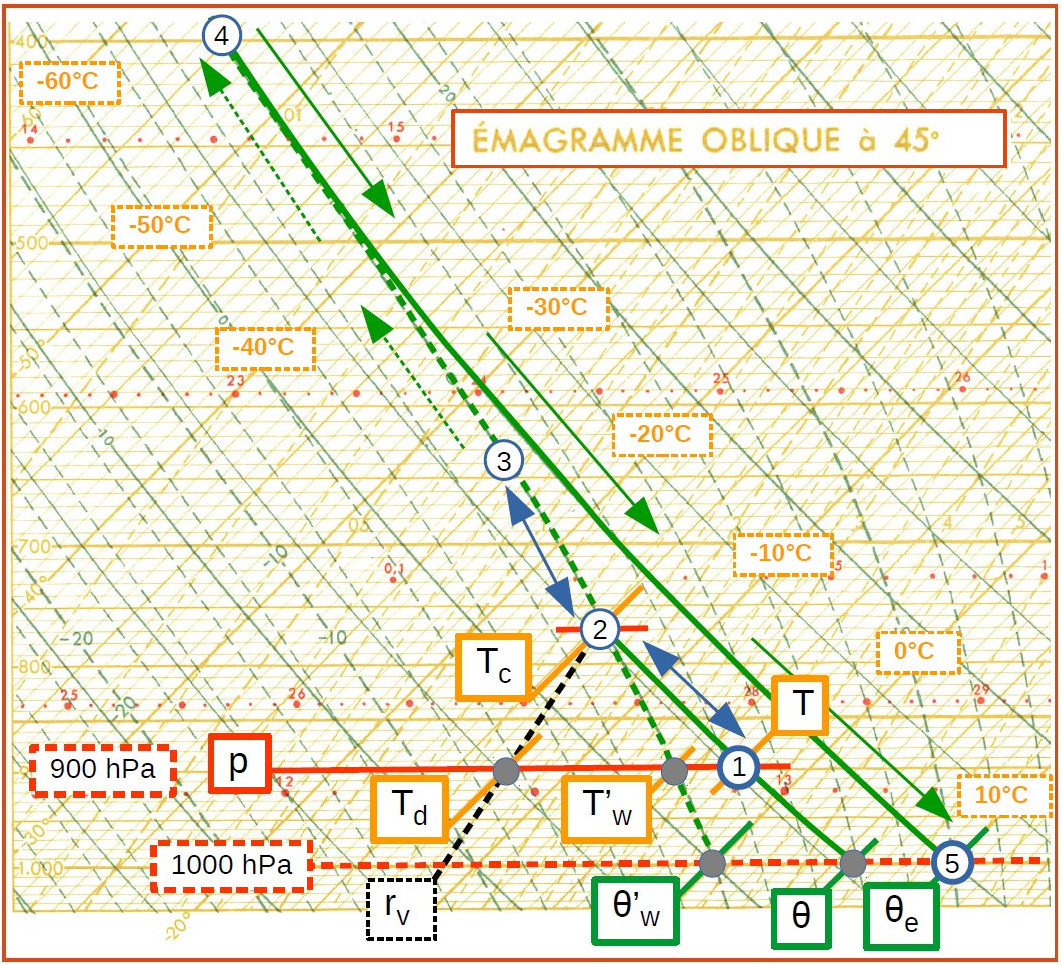}
\vspace*{-2mm}
\caption{
{\it
The different temperatures and potential temperatures studied in this article are plotted on the slantwise Emmagram (the one used in France).
The pressure is in ordinates and the temperature in abscissas. 
The maroon slantwise lines represent the temperature ($T$, solid) and the water vapour mixing ratio ($r_v$, dashed).
The horizontal maroon lines represent the pressure.
The green lines represent the (dry-air) adiabatic potential temperature $\theta$ (solid) and (saturated moist-air) pseudo-adiabatic potential temperature $\theta'_w$ (dashed). 
The condensation, dew-point and wet-bulb temperatures are $T_c$, $T_d$ and $T'_w$, respectively.
}
\label{Fig_emagramme}}
\end{figure}

The adiabatic law is represented on the Emmagram on the figure~\ref{Fig_emagramme} by the continuous green lines, with a conservation of this quantity ``$\theta$'' between the two points 1 and 2.

Without reference to the work of Poisson, the English Joule (1845) and Thomson (1862, the future Lord Kelvin) searched for the laws that describe the temperature changes associated with compression and expansion of gases, as well as the impact due to the possible condensation of water.

Thomson used the law ``$ T \: p^{(1- \gamma) / \gamma} = Cste $'' to deduce that the temperature should vanish at the height of $30$~km, where adiabatic movements  prevail in an hydrostatic equilibrium atmosphere (the state of convective equilibrium state).
Thomson believed this impossible, and in addition to imagining the possible impact of radiation to prevent the temperature from becoming negative (to become a ``radiative-convective equilibrium'' state), he evaluated the effect previously predicted by Joule due to the condensation of water vapour in cloudy saturated areas.
This was the first assessment of the impact of the saturated moist-air adiabatic gradient, which is in the range of $-0.6 $~K per $100$~m, a value actually much lower than the dry or unsaturated air value of about $ -0.9 $~K per 100~m.
These pseudo-adiabatic motions are close to those represented by the dotted green curves in Figure~1, with a vertical gradient of temperature actually less important (in absolute values) between points 2, 3 and 4 than between points 1 and 2.

\clearpage

After the definition of the entropy function by Clausius in 1865, his ideas were quickly used in Germany in the papers of Hertz (1884), von Helmholtz (1888) and finally von Bezold (1888a), where his equations~9 and 10 (p 504) represent the differential of adiabatic and pseudo-adiabatic movements of moist air. 
It is indeed in this same page 504 that von Bezold defines for the first time the notion of ``Pseudoadiabate'' (pseudo-adiabatic in English), when the condensates are eliminated in the form of precipitations.
He started from the differential equation for the entropy and he derived the differential equation for the quantity now denoted $\theta'_w$ and which is conserved between the points 2, 3, 4 in the Emmagram on the figure~\ref{Fig_emagramme}, where the green dashed lines are those where the wet-bulb pseudo-adiabatic potential temperature $\theta'_w$ is conserved.
It is the impact of the entropy withdrawn by the precipitations (or due to the water vapour added to maintain the saturation) that explains the difference between the solid ($\theta$) and dashed ($\theta'_w$) green lines.
These papers of 1888 are specifically concerned with the thermodynamic properties that prevail during the movements of the atmosphere, and to facilitate the developments of meteorology (Entwickelung der Meteorologie).

The Greek letter ``$\theta $'' is first used by von Helmholtz (1888) to represent the absolute temperature (p.650), before to denote (p.652) more specifically the temperature than a mass of air would acquire if it were adiabatically moved to a given standard pressure $p_0$ following the Thomson's law `` $ \theta \: {p_0}^{(1- \gamma) / \gamma} = T \: p^{( 1- \gamma) / \gamma} $'', the law deduced from the Poisson equations. 
This transformation is shown in Figure~1 to go from point 1 to that of temperature $ \theta $ at the pressure of 1000~hPa. 
Helmholtz first called this quantity ``Wärmegehalt'' (total content in heat), before von Bezold (1888b, p 1189) proposed, with the explicit agreement of Helmholtz, the name of ``potential temperature'', the name that has lasted until today.

It must be emphasized that the link between the potential temperature of the dry air and entropy was not made by von Helmholtz nor by von Bezold, except through a small remark from the latter (1888b, page~1193) where it is said that the properties of $ \theta $ ``resemble those due to Clausius' theorem, while being different from this theorem''. Here, reference is made to the second principle of thermodynamics and to the fact that entropy ``tends to a maximum'', in the same way as von Bezold showed that $ \theta $ is ``preserved by adiabatic movements of moist air in the free atmosphere (adiabatic and isentropic transformations), or can only increase in the presence of condensations removed by precipitations'' (pseudo-adiabatic transformations).

Based on this remark made by von Bezold in 1888, it was not until Bauer's paper (1908) that the link between the entropy of dry air ($s$) and the potential temperature ($\theta$) was clearly established, in the form\::
\vspace{-1mm}
\begin{align}  
 s & \: = \; c_{pd} \: \log\left(\theta\right) 
 \: + \:  S_1 \; .
 \label{eq_1}
\end{align}

This law explains that, to calculate the entropy of the dry air (up to an additive constant ``$S_1$''), one must take the logarithm of $ \theta = T \: (p_0 / p)^{ \, \kappa}$ with $\kappa = R_d/c_{pd} \approx 0.2857 $ and $p_0 = 1000$~hPa, then multiply the result by the specific heat at constant pressure $c_{pd}$ which is a constant close to $1005$~J/K/kg for the atmospheric range of temperature. 
We obtain then the good law of variation of the entropy of a perfect gas according to logarithms of its temperature and its pressure:
\vspace{-1mm}
\begin{align} 
 s (T , p) & \: = \; 
 c_{pd} \: \log\left( T \right) 
 \: - \:
 R_d \: \log\left( p \right) 
 \: + \:
 S_2 \; ,
 \label{eq_2}
\end{align}
where $ R_d \approx 287$~J/K/kg and where the constant $S_2$ is different from $S_1$ (due to $p_0$).

But these relations (\ref{eq_1}) and (\ref{eq_2}) obtained in 1908 are only relevant for dry air, without giving any indication of a possible definition of the ``potential temperature of the moist air'' in connection with its entropy, even though this quantity can exist.

 \section{\underline{\Large The equivalent potential temperatures}}
\label{section_2}
\vspace{-1mm}

It is often considered that the answer to this question is the so-called ``equivalent'' potential temperature and denoted ``$\theta_e $'', that it would be enough to put in the logarithm of (\ref{eq_1}) to give the value of the Entropy of moist air\,? 
In fact it is not, and the title of this section is plural to remind that there are many ways to understand this meaning ``equivalent'' in meteorology. 
Moreover, this plurality is impossible because entropy is a state function in thermodynamics and must lead to unequivocal answers as to its variations between two instants, or between two points.
These remarks require us first and foremost to ask the question: ``equivalent'' to what\,?

The answer is given by two students who were in thesis with von Bezold: Schubert (1904) and Knocke (1906). They called ``äquivalente'' or ``ergänzte Temperatur'' the one ``completed'' by the ``supplement'' of the energy due to the water content. 
The idea is to add to the total energy (in fact the enthalpy of the dry air measured by the product ``$ c_{pd} \: T $'') the ``energy'' due to the latent heat release, which is measured by ``$ L_v \: q_v $''. 
They thus form the sum $c_{pd} \: T_e = c_{pd} \: T + L_v \: q_v$, where $T_e$ is by definition the equivalent temperature, and where the enthalpy of the dry air is therefore increased by the product of the latent heat of vaporization ($L_v$) and the water vapor mass content ($q_v$).
Schubert and Knocke have retained the suggestion of von Bezold to use the name ``equivalent temperature'' to denote\,: $T_e = T + L_v(T) \: q_v / c_{pd}$, where the values of $L_v(T)$ depend on the absolute temperature.

Next, Normand (1921) started from Bauer's conclusions that the relations (\ref{eq_1}) and (\ref{eq_2}) represent the entropy of dry air, but are not applicable to the moist air. 
Normand was able to establish approximated formulas for the pseudo-adiabatic ($\theta'_w$) and equivalent ($\theta_e$) potential temperatures, considering that they are both (thus at the same time\ldots because not so different from each other\ldots) measurements of the entropy of the moist air (even if it is impossible). 
With modern notation, Normand obtained the following approximate relationships for specific (i.e. per unit mass of moist air) moist-air entropy:
\vspace{-1mm}
\begin{align} 
 s  & \: \approx \; 
 c_{pm} \: \log\left( \theta_e \right) 
 \: + \:  s_m \; , 
 \label{eq_3}
 \\
 \mbox{with~:} \; \; \;
 \theta_e  & \: \approx \; 
 \theta \: \left[ 1 \: + \: 
        \frac{L_v \: q_v }
        {c_{pd} \:  T} \right] \; , 
 \label{eq_4}
 \\
 \mbox{or, similarly~:} \; \; \;
 \theta_e  & \: \approx \; 
   \frac{\theta}{T}
   \; \left[ T \: + \: 
        \frac{L_v \: q_v }
        {c_{pd}} \right] 
 \label{eq_5}
  \; .
\end{align}
The two formulations (\ref{eq_4}) and (\ref{eq_5}) are similar, but we recognize more clearly in (\ref{eq_5}) the factor $L_v \: q_v / c_ {pd}$ introduced by von Bezold, Schubert and Knocke (all 3 cited by Normand).
This factor explains the English name ``equivalent'' used by Normand to qualify the potential temperature ``$\theta_e \approx (T_e / T) \: \theta$'', where $T_e = T + L_v \: q_v / c_{pd}$ has the same definition as in Schubert and Knocke (under the approximation $ q_v \approx r_v $ for the specific content and the mixing ratio, which is made throughout the present paper).

But the interest of (\ref{eq_3}) and of the definition (\ref{eq_4}) is weakened by the approximations made by Normand to integrate (in the mathematical sense of the term) the von Bezold differential equation for the entropy of moist air. 
In particular, Normand did not apply the third principle of thermodynamics described in Part~1, assuming arbitrarily that the entropy of $1$~kg of saturated air plus $14.7$~gr (?) of liquid water can be set to zero at $0$~Celsius. 
The third principle expresses, differently, that the entropies must be set to zero at $0$~Kelvin (or $-273.15$~Celsius) only for the {\it most stable solid phases\/} of all the components of the atmosphere (N${}_{2}$, O${}_{2}$, H${}_{2}$O, Ar, CO${}_{2}$, etc).
The consequence is that the term $s_m$ in (\ref{eq_3}) is not the one required by the third-law of thermodynamics, with a missing and variable term in $s_m$ which depends on the total water content $q_t=q_v+q_l+q_i$ (water vapour plus liquid water plus ice) and which explains the larger part of the approximation ($\approx$) in (\ref{eq_3}).

Moreover, both $c_{pm}$ and $s_m$ in (\ref{eq_3}) depend on the total water content ($q_t$), so that $\theta_e$ cannot vary in the same way as the entropy ``$s$'' if ``$q_t$'' is variable in space and time (which is the case everywhere in the atmosphere). 
Thus, the ``equivalent'' aspect of $\theta_e$ with entropy would be based on the fact that the multiplicative ($c_{pm}$) and additive ($s_m$) factors relative to the logarithm may be constant.
But this is not true, and differently the aim of Normand was to find a certain link between the entropy in terms of the same ``equivalent'' temperature $T_e = (\theta_e / \theta) \: T$ of von Bezold, Schubert and Knocke, whatever $c_{pm}$ and $s_m$ might be variables (impact of $q_t$) and inaccurate (different from third-law values).

There are two ways to illustrate that the variable $\theta_e$ can not represent the entropy of moist air, in the most general case where the water content is variable in time and space.

The first approach is to imagine an isentropic region, where the specific entropy ``$s$'' is constant in (\ref{eq_3}). 
In this case the reciprocal formula
\vspace{-1mm}
\begin{align}
 \theta_e  & \: \approx \; 
 \exp \left( \frac{s \: - \: s_m}{c_{pm}} \right)
 \label{eq_3_reverse}
  \;
\end{align}
leads to a priori variable values for $\theta_e$ even though $s$ is constant, since both $s_m$ and $c_{pm}$ are variable in the argument of the exponential function.
Moreover, the fact that the term $s_m$ is not the one expected by thermodynamics amounts to saying that other variable terms must be added or subtracted in $s_m$, making the links (\ref{eq_3}) and (\ref{eq_3_reverse}) between the specific entropy ``$s$'' of Normand and $\theta_e$ more inaccurate.
On the other hand, the numerical results described in next Chapters~3 and 4 invalidate the possibility that the impacts of joint variations of $s_m$ and $c_{pm}$ can compensate each others to give a constant value for $\theta_e$ in (\ref{eq_3_reverse}).

The second approach is to imagine the case where $\theta_e$ is constant in a region, on a surface or in a given line. But for the case where the total water content is not constant and where both $s_m$ and $c_{pm}$ are variable, then according to (3) the specific entropy $s$ must become variable, which prevents imagining direct and universal links between the Norman formula for $s$ and $\theta_e$, on the one hand, and with the entropy of moist air defined according to the third law and the precepts of thermodynamics, on the other hand.

These major defects are found in all formulations of $\theta_e$, in spite of all the care taken to look for a quantity that can be equivalent to the entropy of moist air.
Indeed, from Normand (1921) the meaning of the word ``equivalent'' was gradually transformed, this time with a desire to find an equivalent to the entropy of moist air, and no longer to the {\it impact of moisture on the energy and enthalpy\/}, as originally planned by von Bezold, Schubert and Knocke.

Several other equivalent potential temperature formulations have been defined, but they all correspond, in good approximation, to the approximate links discovered by Normand in 1921 between the entropy of moist air and $\theta_e$, links that are given by (\ref{eq_3}) and (\ref{eq_3_reverse}).

Like Normand, the study of Rossby (1932) is based on the aerological vision and the emagram which means that the equivalent temperature ($\theta_e$) is obtained by rising toward infinite heights (the point 4 on figure~1) following a pseudo-adiabatic (iso-$\theta'_w$), then going down to $1000$~hPa according to a dry adiabatic (iso-$\theta$) up to point 5 of temperature $\theta_e$. 
By making different approximations, Rossby found almost the same formulations (\ref{eq_3}) and (\ref{eq_4}) of Normand. 
But for this purpose, Rossby also used moist values for $c_{pm}$ and $s_m$ which are not constants and prevents $\theta_e$ from being synonymous with entropy. 
And anyway, Rossby used as Normand pseudo-adiabatic transformations that are not isentropes, because of the irreversible nature of the elimination of condensates by precipitation. 
The links suggested by Normand and Rossby between $\theta_e$ and the entropy of moist air can therefore only be approximate and they do not correspond to the definition of entropy given by the thermodynamic and the third law.

In the more recent article by Betts (1973) the relevant differential equations of moist entropy defined by von Bezold (1888a) and Saunders (1957) are used, but with many approximations further. 
In particular, Betts assumes constant (conservative) the total proportion of water, that is to say the sum of the contents of water vapour and liquid water ``$q_t=q_v+q_l$'', and he also assumes that $ R / c_p \approx R_d / c_ {pd} $ and $ L_v (T) / T \approx L_v (T_0) / T_0 $ are constants.
It is only with all these hypotheses that he was able to define the ``potential liquid temperature'' by removing (arbitrarily) the quantity ``$ L_v \: q_t / (c_ {pd} \: T) $'' in the bracketed factor of (\ref{eq_4}) to get $\theta_e$ from $\theta_l$, leading to:
\vspace{-1mm}
\begin{align} 
 \theta_l  & \: \approx \; 
 \theta_e  \: \left[ 1 \: - \: 
        \frac{L_v \: q_t }
        {c_{pd} \:  T} \right] \: , 
 \label{eq_6}
  \\
 \theta_l  & \: \approx \; 
 \theta  \: \left[ 1 \: - \: 
        \frac{L_v \: q_l }
        {c_{pd} \:  T} \right] 
      \: .
 \label{eq_7}
\end{align}
This variable $\theta_l$ is important because, associated with the total water content ``$q_t$'', they form the pair of the so-called ``conservative'' variables on which turbulence acts in almost all NWP models and GCMs (in ARPEGE, AROME, Meso-NH and LMDZ in France, in IFS at ECWWF, in ALARO at LACE, in COSMO and ICON at the DWD, in the Unified Model at the DWD, \ldots).

It may be useful to clarify the meaning of this term ``conservative''. 
It is not a question of considering that $\theta_l$ and $q_t$ are conserved (constant) everywhere and at any moment. Here we consider ``principles of conservation'' similar to that for energy, where if the energy is indeed conserved for an isolated system, it can grow or decrease depending on the source terms or energy sink. 
Here too, the total water content $q_t = q_v + q_l$ is preserved in the event of phase change and reversible drop creation in a cloud (creation of $q_l$ at the expense of $q_v$) or in the event of evaporation of cloudy drops (creation $q_v$ at the expense of $q_l$). 
This is the application of the principle of ``conservation of matter''. 
But the variable $q_t$ can decrease in the event of precipitation fall (decrease of $q_l$ with $q_v$ unchanged). 
Similarly, according to studies by Normand, Rossby and Betts, the main principle associated with the ``conservative'' aspect for $\theta_l$ or $\theta_e$ seems to be the second principle of thermodynamics and the adiabatic or pseudo-adiabatic equations for entropy.

Emanuel (1994) derived a formulation similar to (\ref{eq_4}), but without going through the integration of the differential equation of entropy of moist air, as Rossby and Betts did.
In the same way as Normand before, Emanuel's approach is based on a direct calculation of the entropy of moist air, as a weighted sum of the entropy of its constituents. 
However, in a way similar to what Betts did, Emanuel have added, subtracted and multiplied several ``conservative'' quantities depending on ``$q_t = q_v + q_l$'', which are arbitrarily considered constant.
We thus find the same problems related to the appearance of the variable quantities $c_{pm}$ and $s_m$ in (\ref{eq_4}), which both prevent the variable $\theta_e$ defined by Emanuel from being synonymous with entropy in all circumstances.
Moreover, to carry out his calculations, Emanuel does not apply the third principle of thermodynamics (in the same way as Norman did before him) by cancelling entropies at $0$~Celsius, and not at $0$~Kelvin, this leading to missing terms in $s_m$.

The many and more recent approaches of Pauluis et al. (2010), Pauluis et al. (2011), Mrowiec et al. (2016) are based on the same technique as that of Emanuel (the moist-air entropy is defined by the weighted sum of those of its constituents), with the same presence of variable quantities $c_{pm}$ and $s_m$ in (\ref{eq_4}), and without applying the third principle for the atmosphere. 
We can note in addition the confusion made by Pauluis when he wrote in 2010 that the term in ``$\ln (T)$'' in (\ref{eq_2}) becomes infinite when $T$ goes to absolute zero, which in his opinion prevents the application of the third principle to the  moist-air atmosphere. 
This argument is misleading because, as explained in the first part of the article, the third principle only applies to solid phases close to $0$~K, and no to gases, and with a finite variation of entropy as a function of temperature.

Thus, none of the approaches of Normand, Rossby, Betts, Emanuel or Pauluis allowed to find through $\theta_e$ or $\theta_l$ a potential temperature which is synonymous with the entropy of the moist air in general, and for all thermodynamic conditions. 
We could therefore ask ourselves the questions: should we continue this quest for a ``potential entropic temperature''\,? Was this dream even realizable\,?

------------------------------------------

 \section{\underline{\Large The absolute entropy in meteorology}}
\label{section_3}
\vspace{-1mm}

In response to the first question, we can cite the need to use the entropy of moist air in the equations of meteorology which was explicitly indicated by Bjerknes (1904, 1995), where the equation resulting from the second principle of thermodynamics is set aside only for the simplest case where the water content is a constant (as for the definition of $\theta_e$ and $\theta_l$ by Betts, Emanuel and Pauluis).

As for the pursuit of Bjerknes' quest, Richardson (1922) soon afterwards worked out the entropy of moist air, already mentioning in his book that the criticism of infinity of ``$\ln(T)$'' in (\ref{eq_2}) when $T$ tends to $0$~K is a false problem, and that Nernst's theorem should be applied to the solid phases at $0$~K.
Indeed, Richardson already knew that calorific capacities $c_p(T)$ vary like the cube of the temperature (see the first part of the paper), leading to finite values for the integrals of $c_p(T)/T$. 
But the absence of precise measurements of the entropies at the beginning of the 20th century for all the different gases prevented Richardson from computing values of entropies of all components of the atmosphere\,: N${}_{2}$, O${}_{2}$, H${}_{2}$O, Ar, CO${}_{2}$, ...

The first evaluation of the thermodynamics moist-air entropy to meteorology have been published by Hauf and Höller (1987), with a use of the third-law reference values. 
However, the potential temperature denoted by $\theta_S$ defined by Hauf and Höller is related to the moist entropy by relations similar to (\ref{eq_3}) and (\ref{eq_3_reverse}), with values of $c_{pm}$ and $s_m$ which are not fixed and which vary with the humidity.
Therefore, as indicated in Chapter~2, the presence of $q_t$ outside of the logarithm in $s = c_{pm}(q_t) \: \log(\theta_S) \: + \: s_m(q_t)$ prevents the potential temperature $\theta_S$ of Hauf and Höller from being synonymous with the entropy of moist air, like the variable $\theta_e$ of Normand, Rossby, Betts and Emanuel.
The only way to define a potential temperature that is synonymous with entropy is to include all the parts that depend on the variable $q_t$ in the logarithm, and to apply the third principle of thermodynamics in order to get a relevant value for $s_m$.

In order to solve these issues, I have used in Marquet (2011) the same definition for the moist-air entropy ($s$) as that of Hauf and Höller, but with the will to solve the problem of $c_{pm}$ and $s_m$ variables in (\ref{eq_3}), because depending on the humidity. 
I have been able to write 
\vspace{-1mm}
\begin{align} 
 s  & \: = \; 
 c_{pd} \: \log\left( \theta_s \right) 
 \: + \:
 s_0 \; , 
 \label{eq_8}
\end{align}
to define a new potential temperature $\theta_s$ which depends (as a first approximation) on the two Betts variables ($\theta_l, q_t$) according to:
\vspace{-1mm}
\begin{align} 
 \theta_s  & \: \approx \; 
 \theta_l \: \left( 1 \: + \: 
        5.87 \: q_t \right)
 \label{eq_9}
 \; .
\end{align}
It is an exact equality in (\ref{eq_8}), where both $c_{pd}$ and $s_0$ are true constants.
Here lies the main improvement over the Hauf and Höller's formulation (\ref{eq_3}). 
The approximation in (\ref{eq_9}) is not due to assumptions to define and compute $\theta_s$, but to simplifications made to write this article and to have a formulation that is simple and close to (\ref{eq_4}). 
The exact and complete formulations of $\theta_s$ exist and are given in Marquet (2011, 2017).

The coefficient $5.87$ is close to $6$ and is a direct consequence of the third law of thermodynamics.
This coefficient depends on the entropy for solid phases at $0$~K for N${}_{2}$, O${}_{2}$, H${}_{2}$O, Ar, CO${}_{2}$, ... 
This coefficient is about the $2/3$ of the coefficient ``$ L_v / c_{pd} \: T \approx 9 $'' which is involved in (\ref{eq_4}) to (\ref{eq_7}). 
Equations (\ref{eq_6}) and (\ref{eq_9}) imply 
$\theta_s \approx \theta_l \: + \: 2/3 \: (\theta_e-\theta_l)$ and we can thus expect $\theta_s$ to be in the $2/3$ position versus $1/3$ between $\theta_l$ (coefficient $0$) and $\theta_e$ (coefficient $9$).

\begin{figure}[hbt]
\centering
\includegraphics[width=0.68\linewidth,angle=0,clip=true]{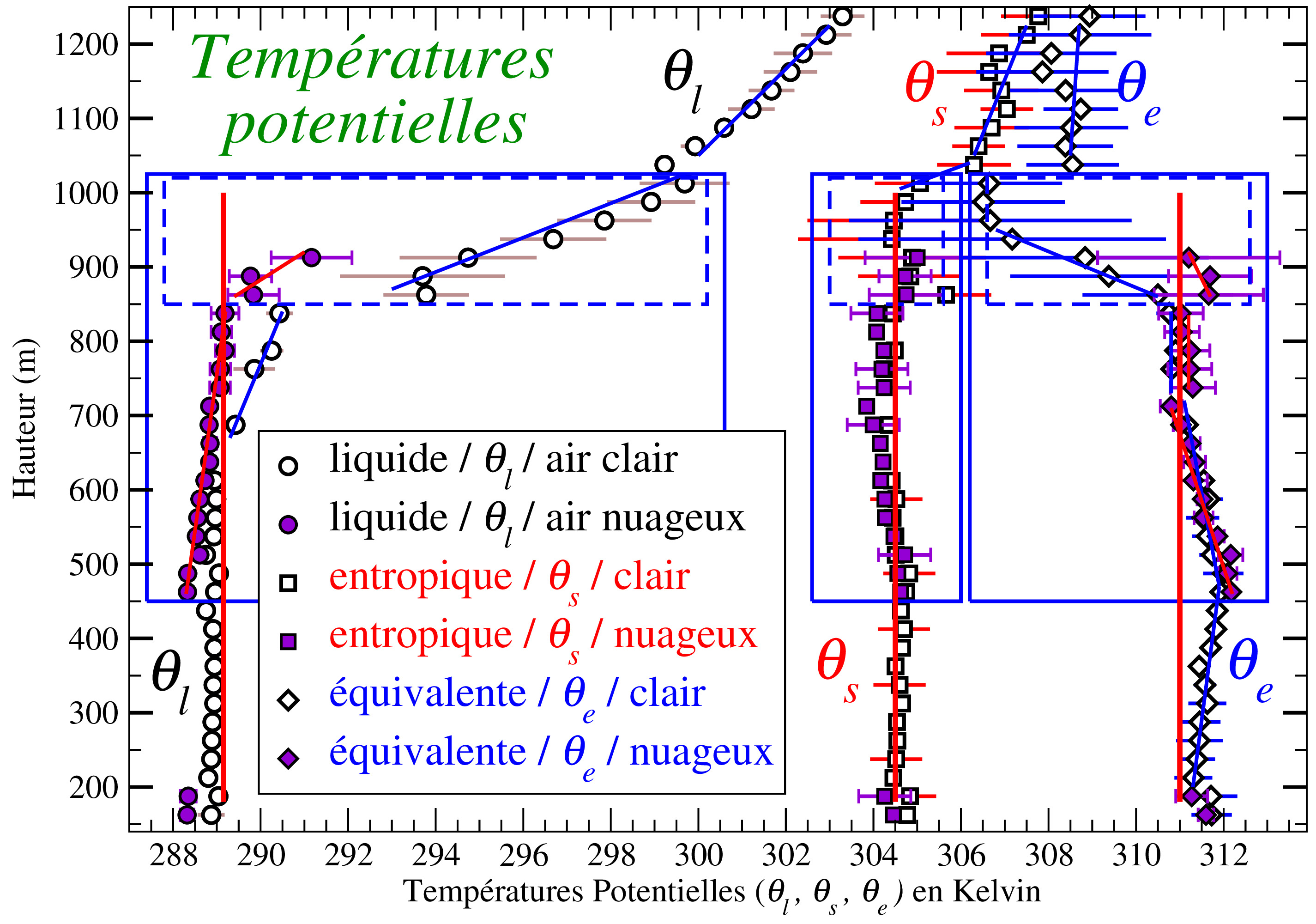}
\includegraphics[width=0.31\linewidth,angle=0,clip=true]{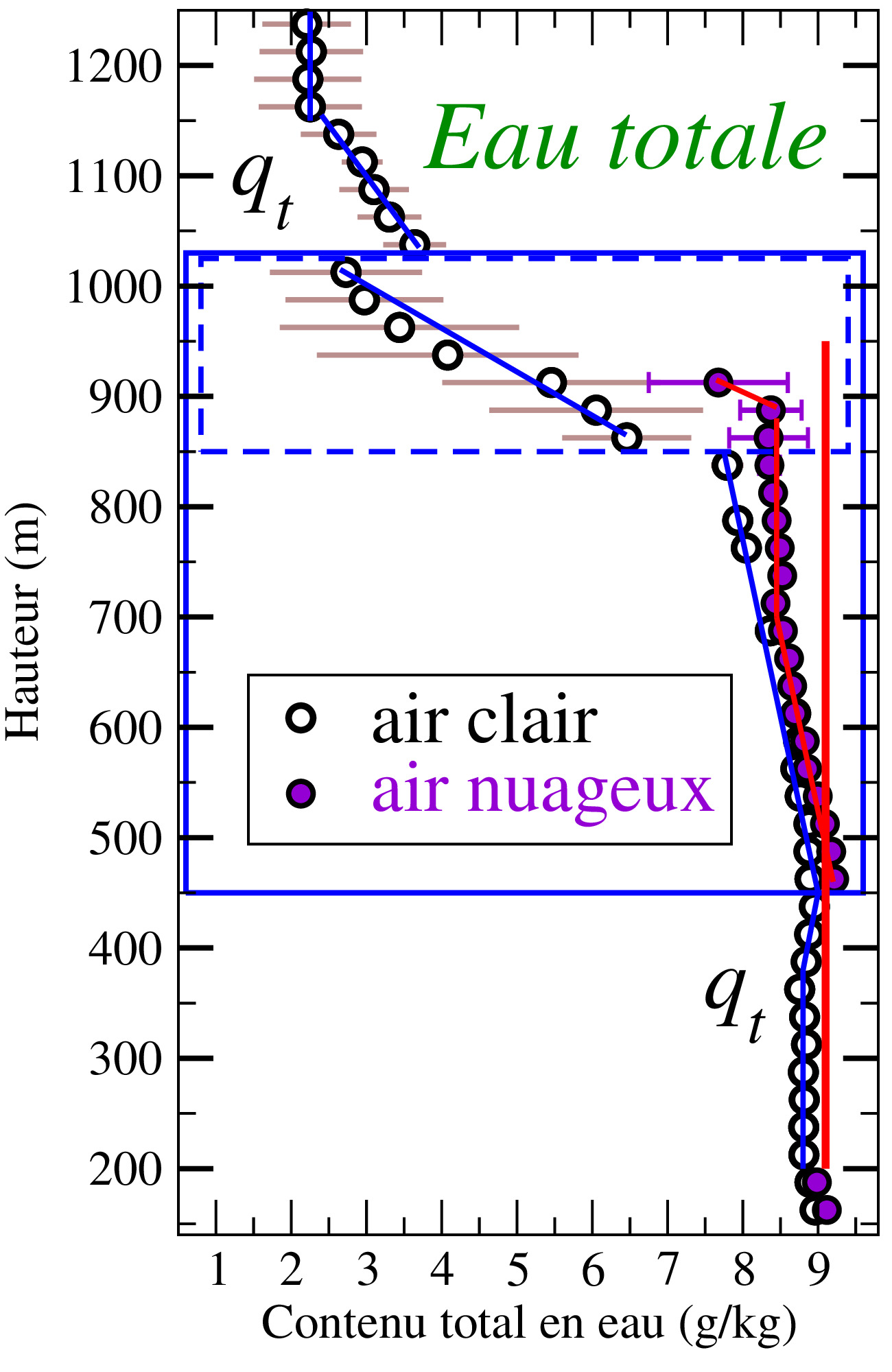}
\vspace*{-3mm}
\caption{
{\it
In this figure adapted from Marquet (2011) are represented for the FIRE-I campaign (flight number 3), from left to right: the vertical profiles of the 3 potential temperatures $\theta_l$, $\theta_s$ and $\theta_e$, as well as the total content of water $q_t = q_v+q_l+q_i$.
Dark purple symbols represent data in cloudy air, while symbols filled in white represent data in clear air regions.
Horizontal error bars represent the standard deviations due to the variability of temperature and humidity measurements.
The rectangles in solid blue represent the stratocumulus located between $450$ and $1025$~m.
The rectangles in dashed blue delimit the entrainment regions between $850$ and $1025$~m, which is the part influenced by the inclusion of drier and warmer air located above.
}
\label{Fig_FIRE_I}}
\end{figure}

This prediction can be validated with the observed data set of the FIRE-I campaign, where several stratocumulus were sampled by several aircraft flights (Marquet, 2011). 
The vertical profiles plotted in Figure~\ref{Fig_FIRE_I} show that, in the boundary layer between $200$ and $800$~m, the values of $\theta_l$, $\theta_s$ and $\theta_e$ are close to $289$, $304$ and $311.5$~K, respectively, with indeed a $(304-289)/(311.5-289)=15/22.5 = 2/3$ position of $\theta_s$ between $\theta_l $ and $\theta_e$.

But beyond this verification, one observes for the entropy and $\theta_s$ singular and very interesting behaviours within the limit of the uncertainties of measurement, which are indicated by errors bars in Figure~\ref{Fig_FIRE_I}.
Indeed, whereas the values of $\theta_l$, $\theta_e$ and $q_t$ vary with altitude, both in the boundary layer under the cloud ($200$ to $450$~m), in the cloud ($450$ to $850$~m) and especially in the entrainment region ($850$ to $1025$~m), the values of $\theta_s$ are almost constant in Figure~\ref{Fig_FIRE_I} to less than $0.5$ degree, included in the entrainment layer where the large increase of $10$~K for $\theta_l$ is almost balanced by the large decrease of $5$~g/kg for $q_t$. 
This is a first remarkable and unexpected result, because nothing imposes a priori that the entropy and $\theta_s$ are constant, and this isentropic region was not visible until now through the plots of the profiles of $\theta_l$ or from $\theta_e$.

Moreover, whereas the values in the cloudy areas and in the clear sky parts are quite different for $\theta_l$, $\theta_e$ and $q_t$ (especially in the entrainment region) they can almost superimpose here for $\theta_s$, and thus for the moist-air entropy (see Marquet, 2011).
This is the second remarkable and unexpected result that neither $\theta_l$ nor $\theta_e$ had revealed so far.

These results are only obtained with this coefficient of $5.87$ which is a direct consequence of the third law, and which explains these mysterious balances that nobody could have guessed a priori.

Since entropy is a state function that depends only on local conditions of temperature, pressure and water content, it cannot ``at the same time'' increase ($\theta_l$), decrease ($\theta_e$) or remain constant ($\theta_s$) with height in the boundary layer and the entrainment region. 
These three variables ($\theta_l$, $\theta_e$, $\theta_s$) cannot be conserved (constants) at the same time: at most one of them can represent the entropy of the moist air. 
It turns out to be $\theta_s$, because it is connected to entropy by the constant coefficients $c_{pd}$ and $s_0$ in (\ref{eq_8}), and because $s_0$ is computed from the third law of thermodynamics.
We can conclude that the variations with height of $\theta_l$ and $\theta_e$ and the differences between clear-air and cloudy regions are only artefacts due to the variable coefficients $c_{pm}$ and $s_m$ in (\ref{eq_3}), and due to missing terms in $s_m$.

These singular properties observed for the FIRE-I campaign and for this aircraft flight number 3 are confirmed for the other flights of FIRE-I (see Fig.2 in Marquet, 2011), and also for several other stratocumulus profiles (campaigns ASTEX, EPIC, DYCOMS, see Fig. 12 of Marquet, 2011). 
It is therefore a fairly general property to see entropy and $\theta_s$ well mixed in boundary layers of marine stratocumulus.
These results must not correspond to artefacts in the definition of Hauf and Höller (1987) for the specific entropy ($s$), nor in the definition of $s=s(\theta_s)$ and $\theta_s$ given by (\ref{eq_8}) and (\ref{eq_9}).
These results must match original physical properties that neither $\theta_l$ nor $\theta_e$ have.

It turns out that these physical properties were predicted a long time ago by Richardson (1919) where he explained that the variables on which turbulence must act are the components of the wind, the total water content ($q_t$) and the entropy of the moist air.
More precisely, Richardson (1920) explained that the moist-air entropy should be replaced by  the associated potential temperature, without the non-linear effect of the logarithmic function in the definition (\ref{eq_8}) for $\theta_s$.

This vision of Richardson corresponds to the vertical profiles plotted in Figure~\ref{Fig_FIRE_I} for the marine stratocumulus, where the layer ($200$~m -- $1000$~m) between the surface boundary layer and the top of the cloud is isolated from the free atmosphere above, this boundary layer behaving like a turbulent region where the main effect is of homogenizing the entropy despite the heat fluxes ($T$) and matter flow ($q_t$) imposed close to the surface. 

The theorem established by Richardson in 1920 (p.362) can be summarized as follows\,: ``The average rate at which the internal and gravitational energies are jointly transformed into turbulent energy is proportional to the vertical gradient of entropy''.
In doing so, the fact that the entropy is constant along the vertical in Figure~\ref{Fig_FIRE_I} (zero vertical gradient in $\theta_s$) would indicate a stationary and particular state of zero production of entropy by turbulence.

In this sense, the behaviour for entropy and $\theta_s$ seems different from that for material fluxes ($q_t$), where turbulence cannot homogenize the total water content $q_t$ and with a non-zero vertical gradient for $q_t$.

Here are examples of properties that could not be discovered and highlighted before we can calculate and plot the entropy of moist air via the entropy of Hauf and Höller or the new variable $\theta_s$ (because, differently, $\theta_l$ increases with height and $\theta_e$ decreases with height).

It can be seen in Figure~\ref{Fig_FIRE_I} that the information provided by the vertical profiles of $\theta_l$ and $q_t$ are complementary and exactly opposite. 
Moreover, only $q_t$ has a clear physical meaning, based on the conservation of the composition of the matter, while conversely $\theta_l$ (and $\theta_e$) has less obvious physical meaning. 
As for the entropy, and therefore the variable $\theta_s$, as it differs radically from all the other variables, the fact of being able to calculate it from 2011 open a field of new studies. 
It is thus possible to explore the interest of replacing the pair of variables ($\theta_l$, $q_t$) by ($\theta_s$, $q_t$), where the two variables are derived from general principles of physics\,: the second and third laws of thermodynamics for $\theta_s$; the law of conservation of matter for $q_t$.

\begin{figure}[hbt]
\centering
\includegraphics[width=0.99\linewidth,angle=0,clip=true]{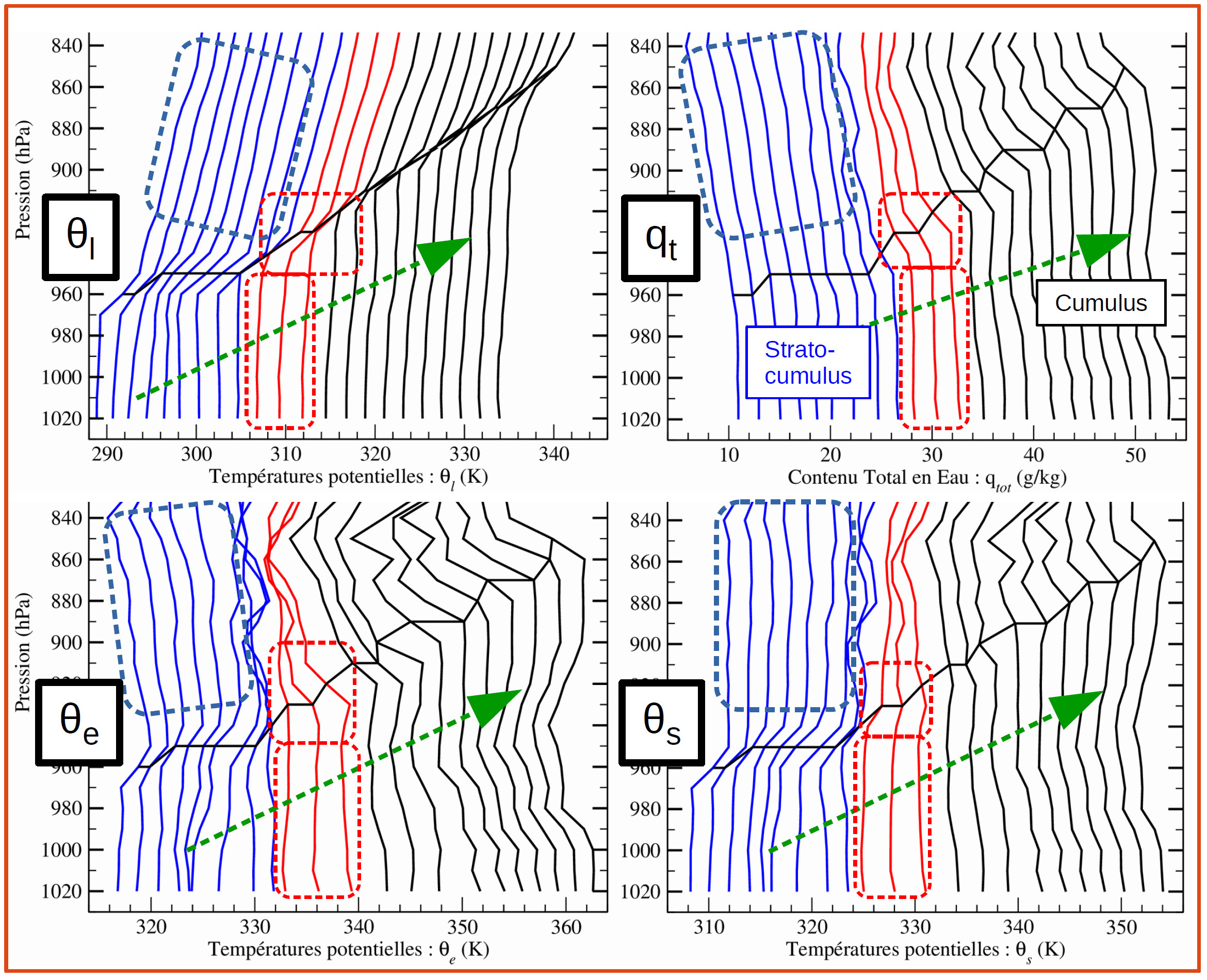}
\vspace*{-3mm}
\caption{
{\it
Vertical profiles corresponding to the aircraft measurements studied in Bretherton and Pincus (1995) for the ``ASTEX Lagrangian'' campaign (digital data available at: \url{https://atmos.washington.edu/~breth/astex/lagr/README.hourly.html}). 
The profiles for stratocumulus are in blue (to the left), those for non-precipitating cumulus are in black (to the right), with in the middle (in red) those for the transition regime. 
For better visibility, only half (one in two) of the 43 aircraft flight are shown, and the successive profiles are shifted by $2$~K for $\theta_l$, $\theta_e$ and $\theta_s$, and $2$~g/kg for $q_t$.
The height of the top of the boundary layer is indicated by the broken line (in black) which is increasing from left to right.
}
\label{Fig_ASTEX_Lag1_hourly}}
\end{figure}

The same results observed for the stratocumulus (FIRE-I, ASTEX, EPIC and DYCOMS campaigns) had to be supplemented by the study of vertical profiles drawn for other types of clouds.
Before showing in the following chapter the results obtained for the more extreme cases of a squall-line and a cyclone, Figure~\ref{Fig_ASTEX_Lag1_hourly} shows the profiles obtained during the ASTEX-Langrangian campaign for non-precipitating cumulus regimes and for the  transitions regimes between stratocumulus and cumulus.
These profiles correspond to Figs.6 of Marquet and Geleyn (2015) and to Figs.37 of Marquet (2016).

We see here in Figure~\ref{Fig_ASTEX_Lag1_hourly} that the vertical profiles are generally more ``regular'', with less ``accident'' for $\theta_s$ than for $\theta_e$.
This means that the combination of $\theta_l$ and $q_t$ is more consistent for $\theta_s$ given by (10), thanks to the constant close to $6$ which comes from the third principle of thermodynamics. 
We also see that the opposite variations of $\theta_s$ and $q_t$ compensate for the regime of stratocumulus (in blue) and lead to an isentropic region where $\theta_s$ is constant above the boundary layer (between $930$ and $840$~hPa).
This is a singular behaviour which is not highlighted with $\theta_l$ or $\theta_e$ which, we still see it here, do not represent the entropy of moist air.

Another remarkable result is that the transition between stratocumulus to cumulus is done for the profiles in red when the jump in entropy (and thus in $\theta_s$) is almost zero at the top of boundary layer.
This transition regime corresponds to a median state between, on the one hand more and more positive jumps (to the left) for stratocumulus, on the other hand more and more negative jumps (to the right) for cumulus clouds.
This result makes it possible to envisage a better analysis of the stratocumulus instability criterion, in connection with the top boundary-layer entrainment processes. 
It should be possible to simplify the current criteria related to the old studies of Randall (1980), Deardorff (1980) or McVean and Mason (1990), where the criteria are based on jumps ``sufficiently positive'' for $\theta_l$, or ``sufficiently negative'' for $\theta_e$, but with very variable and uncertain thresholds for these jumps which are replaced, here, by a simple null jump in $\theta_s$, and with a gain in conceptual simplicity.

As for the cumulus profiles (in black) for $\theta_s$, we see that they look globally similar to those for $\theta_e$, with however for $\theta_s$ less warm values near the surface and a lower decay with the altitude in the lower part of the boundary layer and above its summit. 
This ``non-conservation'' of $\theta_s$ in the boundary layer does not call into question Richardson's principle that turbulence must act on the entropy of moist air.
This indicates that processes related to radiation and associated energy fluxes are becoming more prominent close to the surface due to the opening of the cloud layer, and prevents turbulence from being as effective as for stratocumulus, with vertical gradients existing for both $\theta_s$ and $q_t$.

\clearpage

 \section{\underline{\Large Applications of $\theta_s$ to precipitating systems}}
\label{section_4}
\vspace{-1mm}

Once established in 2011 the formulation (\ref{eq_9}) for $\theta_s$, as well as the first properties of the Hauf and Höller entropy and $\theta_s$ for stratocumulus and non-precipitating cumulus, other aspects were studied in the following years, addressing in particular properties related to saturated and precipitating systems.

The calculation of the Brunt-Väisälä's frequency of moist air has been revisited in Marquet and Geleyn (2013) with the use of the $\theta_s$ variable.
This frequency is an internal variable that is used, via the Richardson number, in some aspects of the turbulence parameterizations.

Potential vorticity calculations have been discussed in Marquet (2014), with the entropy or $\theta_s$ variables that can be integrated into the ``$PV$'' operator, as originally intended by Ertel (1942a, 1942b, see Schubert et al., 2004). 

A summary of these work on the application of absolute entropy to meteorology was written for a chapter of the Convection Book of Plant and Yano (Marquet and Geleyn, 2015). 
Another synthesis, in French, is available in my memoir for the ``Habilitation degree'' (Marquet, 2016).

\begin{figure}[hbt]
\centering
\hspace*{5mm}
{\Large(a)} \includegraphics[width=0.77\linewidth,angle=0,clip=true]{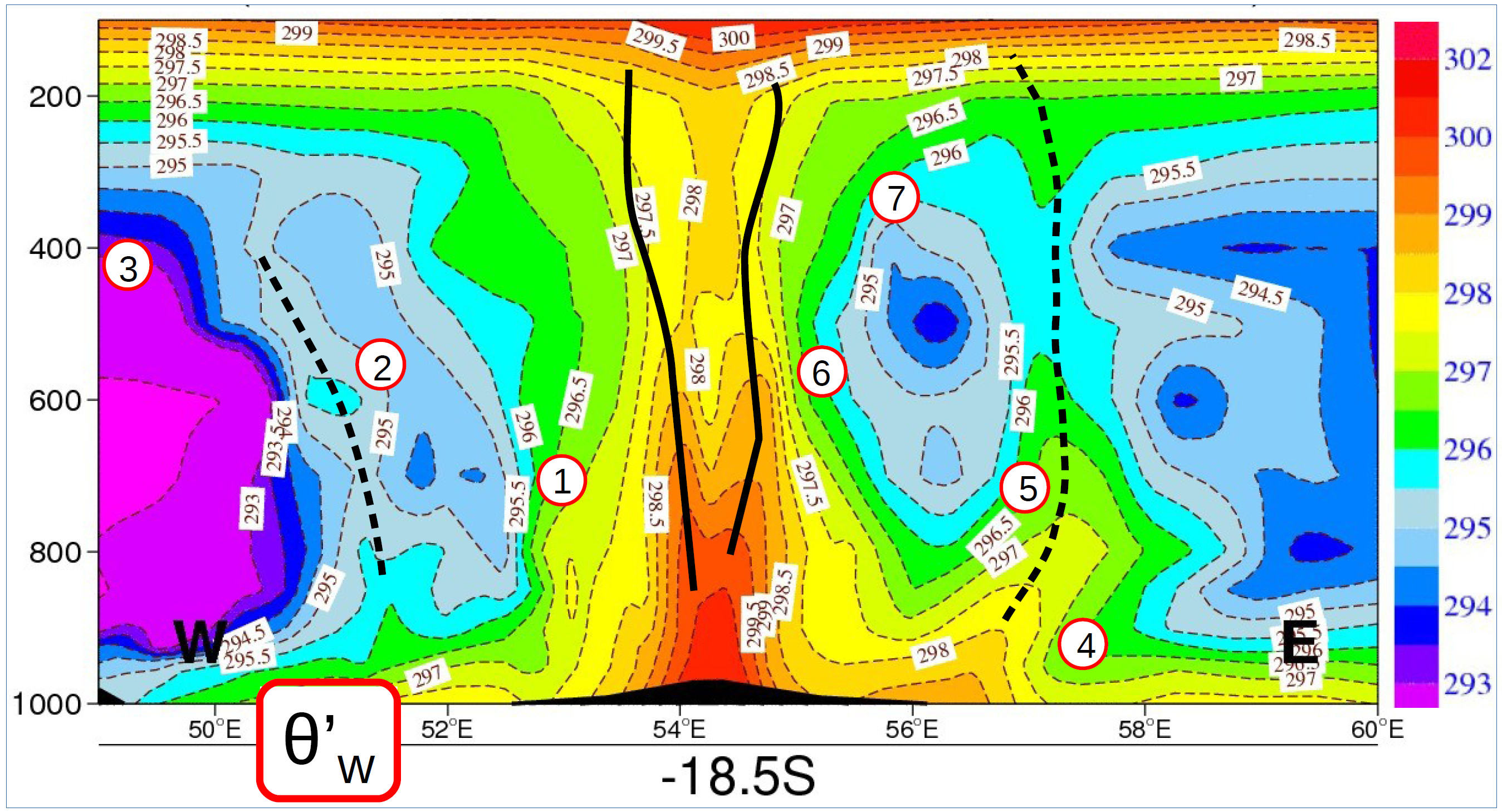}
\\
\hspace{5mm}
{\Large(b)} \includegraphics[width=0.77\linewidth,angle=0,clip=true]{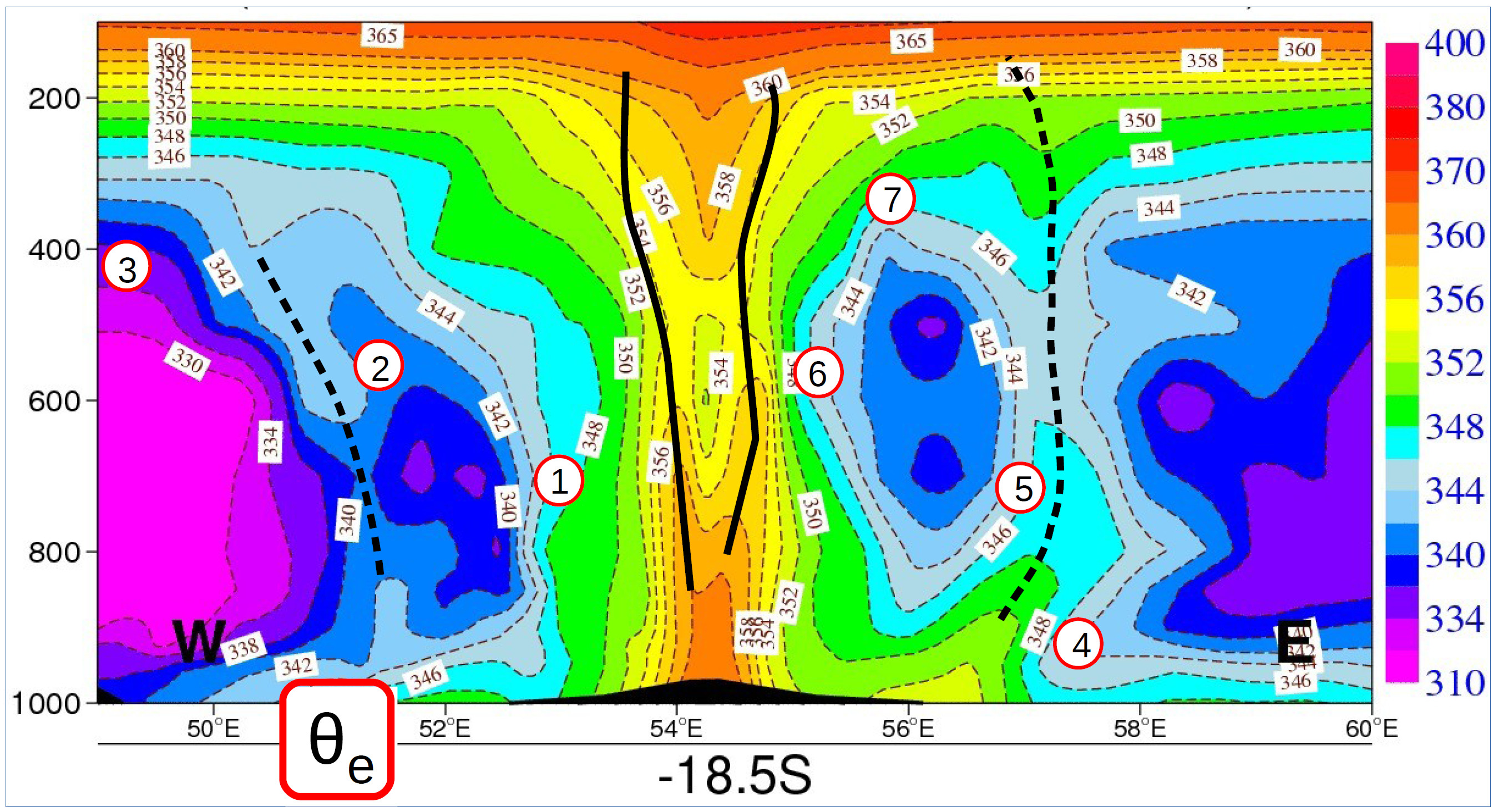}
\\
\hspace{7mm}
{\Large(c)} \includegraphics[width=0.77\linewidth,angle=0,clip=true]{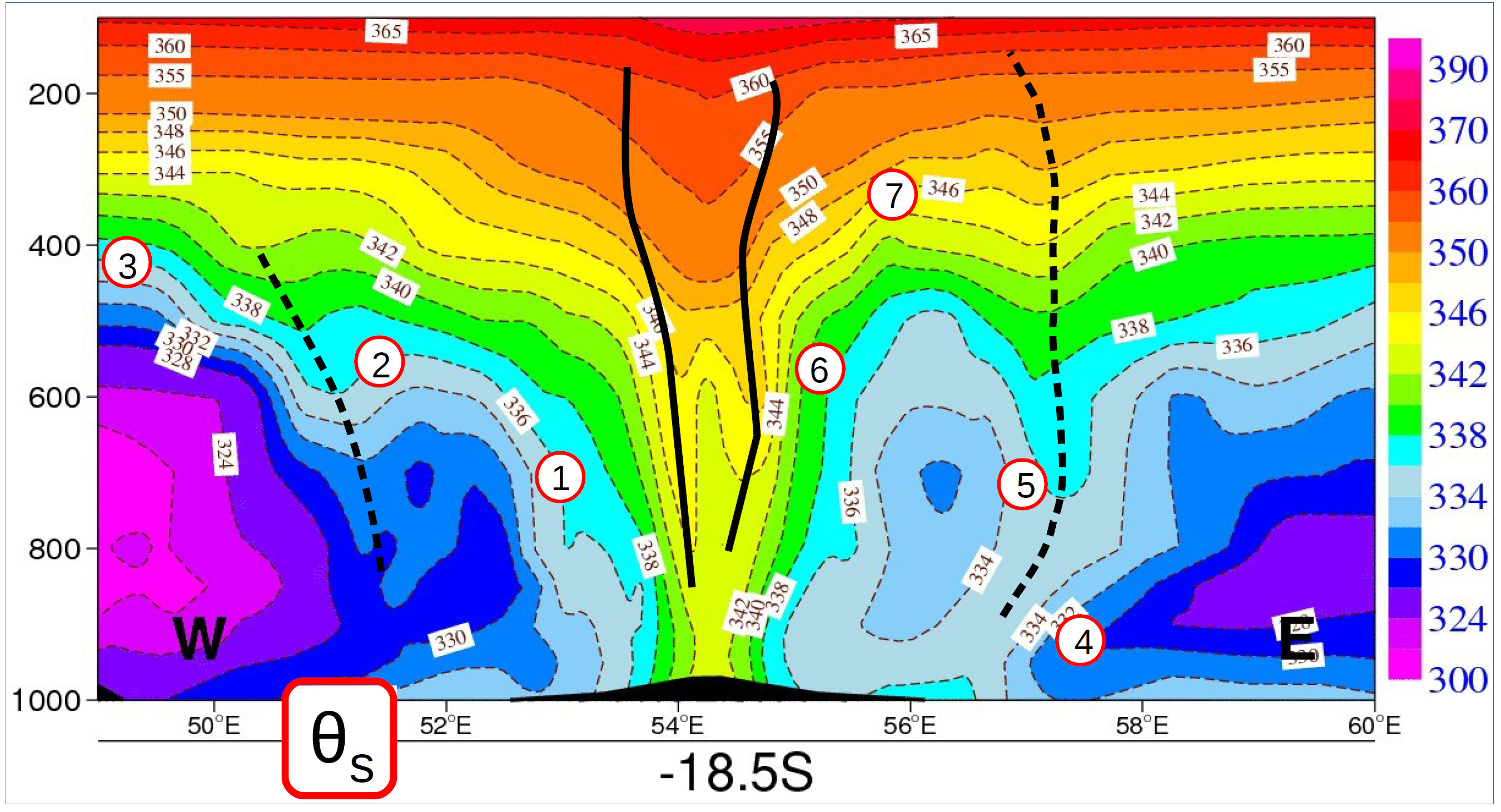}
\vspace*{-1mm}
\caption{
{\it
Vertical sections made in the Cyclone Dumilé (January 3, 2013) for a $12$\,h forecast simulated with the ALADIN-Réunion NWP model.\,: (a) for $\theta'_w$; (b) for $\theta_e$; (c) for $\theta_s$.
The black solid lines represent the eye-wall of the Hurricane and the spiral bands are represented by the dashed black lines. 
These lines have been drawn from saturated areas.
}
\label{Fig_Dumile}}
\end{figure}

More recently, the properties of the moist-air Hauf and Höller's entropy and $\theta_s$ are studied in Marquet (2017) and Marquet and Dauhut (2018) for the cyclone Dumilé (see figure~\ref{Fig_Dumile}) and for the very intense convective system ``Hector the Convector'' (see Dauhut et al., 2018, with Hector corresponding to a very intense convection zone on Tiwi Island in northwestern Australia), both considered as Carnot or thermal machines.
Using the variable $\theta_s$ makes it possible to evaluate the work function and the energy efficiency of these systems, with very different values if we calculate them using the variable $\theta_e$ (Emanuel, Pauluis) or with the true entropy (and $\theta_s$).

We see in figure~\ref{Fig_Dumile} that the patterns for $\theta'_w$ and $\theta_e$ are similar, except the change in color palettes.
This validates the computations of Normand and Rossby for $\theta_e$ as a proxy of a pseudo-adiabatic invariant (and thus being close to $\theta'_w$).

Differently, the visions of the thermodynamic structure of the cyclone are very different if we use $\theta'_w$ and $\theta_e$ on the one hand, or the moist-air entropy and $\theta_s$ on the other hand.
In particular the moist-air isentropes (like between points 1, 2 and 3) can correspond to important differences and strong gradients of $\theta'_w$ and $\theta_e$. 
And conversely, the isolines of $\theta'_w$ and $\theta_e$ indicated between points 4 to 7 does not correspond to an isentrope (iso-$\theta_s$).

In addition, we notice a low-level ``warm core'' in the center of the cyclone with $\theta'_w$ and $\theta_e$, while the moist-air entropy (and $\theta_s$) presents a relative minimum near the ground and increases continuously with altitude. 
This aspect is consistent with the remark made for the cumulus profiles of Figure~\ref{Fig_ASTEX_Lag1_hourly}, where the relative maximum near the surface is more marked with $\theta_e$ than with $\theta_s$.

There are also regions with ``aspirations of moist isentropes'' that correspond to the vision of altitude dynamics, where isentropes (of dry air) dip down from the tropopause to the troposphere.
We can thus extend with $\theta_s$ throughout the troposphere studies of altitude dynamics that are taught from the behaviors of $\theta$, corresponding here to the different convective spiral bands (close to the points 2 and 5).

\begin{figure}[hbt]
\centering
{\Large(a)} \includegraphics[width=0.78\linewidth,angle=0,clip=true]{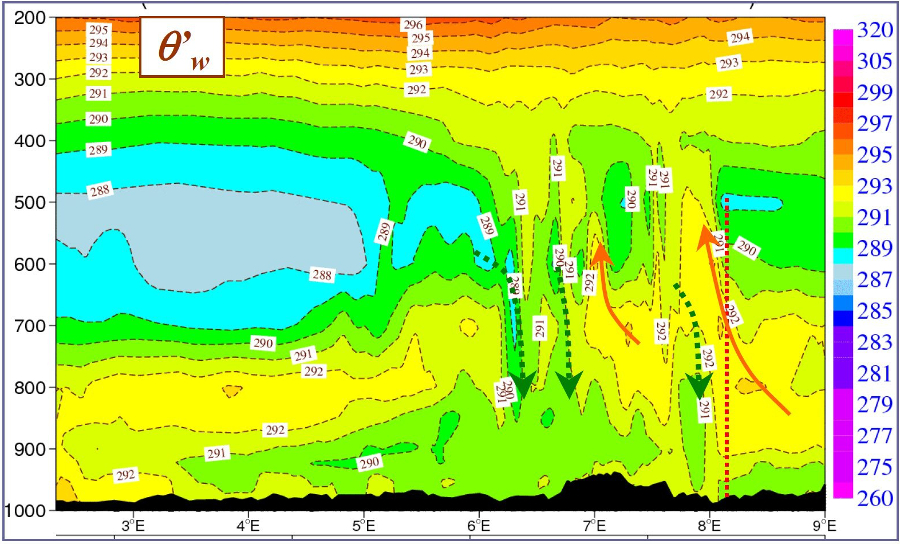}\\
{\Large(b)} \includegraphics[width=0.78\linewidth,angle=0,clip=true]{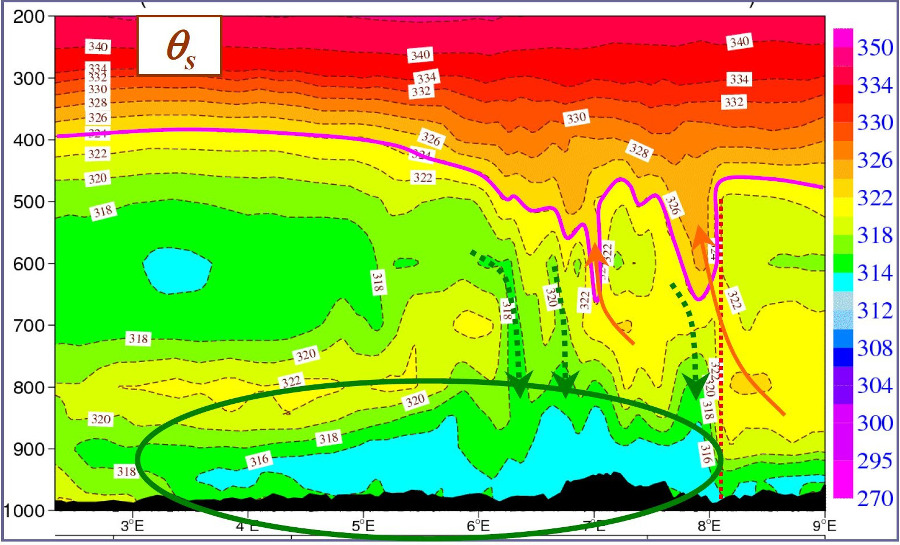}
\vspace*{-1mm}
\caption{
{\it
Study of a squall line simulated by the French AROME NWP model: (a) for $\theta'_w$; (b) for $\theta_s$. 
The front of the squall line is on the right, represented by the  upward orange arrows. 
The back of the squall line is on the left, with the downward green arrows. 
The figures are from Etienne Blot (2013).
}
\label{Fig_Blot_ligne_grains}}
\end{figure}
\clearpage

The use of the moist-air entropy and the variable $\theta_s$ have been evaluated during an internship of Etienne Blot in 2013 (Brochet's Prize 2015).
Many conceptual schemes were created during this internship.
The aim was to describe the interest of the new vision brought by the study of maps and sections made with $\theta_s$ and for the associated potential vortex $PV(\theta_s)$, with the study of fronts, split fronts, convective cells, squall line, hurricanes. 

As for the cyclone Dumilé in Figure~\ref{Fig_Dumile}, the vertical sections of the squall line drawn in Figure~\ref{Fig_Blot_ligne_grains} shows that the vision with $\theta_s$ is very different from that with $\theta'_w$ (and therefore $\theta_e$). 
With the vision in $\theta_s$, the lower value areas (in blue) are better related to the areas of similar valued in the troposphere at the back of the system. 
Above all, we see the strong difference in the impact of the different vertical convective and pseudo-adiabatic regions, where on the one hand $\theta'_w$ is logically constant, while the entropy (and $\theta_s$) logically increases with altitude (creation of entropy by irreversible pseudo-adiabatic processes). 
But we can still clearly identify the convection zones by the ``dipping down of moist-air isentropes and $\theta_s$'', like in Figure~\ref{Fig_Dumile} for the cyclone Dumilé.

 \section{\underline{\Large Conclusions and Outlook}}
\label{section 5}
\vspace{-1mm}

Of course, it may not be a necessity to measure the entropy of the moist air by the value derived by Hauf and Höller, nor with the potential temperature $\theta_s$ given by (\ref{eq_9}).
However, this is possible, and it is shown in this article that new properties can be discovered thanks to this variable $\theta_s$, and thus for the moist-air entropy expressed with reference values of entropies computed from the third law of thermodynamics.

It can be emphasized that the formulation of the moist-air entropy that is taught at the Max Planck Institute (Stevens and Siebesma, 2019) corresponds to the third law reference values and is in full agreement with that of Hauf and Höller (1987), and corresponds to the variable $\theta_s$ too.

This means that in meteorology, in climate as in all fields of physics, it is advisable to apply the precepts of thermodynamics, and therefore here to use the formulation resulting from the third law, which clearly invalidates the use of $\theta_l$ or $\theta_e$ to represent in general the entropy of the moist atmosphere. 

This does not call into question the interest of the variables $\theta_l$ or $\theta_e$ for the special cases of pseudo-adiabatic transformations, or if the composition remains unchanged (constant total water $q_t$), two special cases that enable to make certain links between $\theta_l$ or $\theta_e$ and the moist-air entropy.

But this link is broken as soon as the amount of total water $q_t$ is variable from one point to another and if we draw and compare maps, or calculate gradients or time derivatives. 
And this configuration is the rule in the troposphere everywhere on the globe.

It is now possible to use the potential temperature $\theta_s$ to easily and accurately evaluate the variations of the entropy of moist air in space and time, both numerically and graphically. 
This is a big step forward and an opportunity since 2011 that young scientists should seize to discover new properties of the atmosphere, realizing the dream expressed by Bjerknes in 1904 and by Richardson in 1922.

It must be stressed that the interest of the Hauf and Höller's entropy and $\theta_s$ given by (\ref{eq_9}) is not to find a variable that is necessarily conserved in all cases. 
And besides, she is not. 
The point is that it is a measure of the entropy of the atmosphere under any circumstances, without having to make any assumptions other than those advocated by the third law of thermodynamics and retained in Hauf and Höller (1987).
The study of this variable $\theta_s$ was therefore motivated a priori by general principles of physics, leading to unexpected discoveries of new regions or isentropic surfaces described in the different figures of this article.

The results of these findings should affect all aspects involving moist-air transformation, for example for studies of\,: boundary layer and free atmosphere turbulence, impacts of top-height entraining boundary layer, vertical or slantwise convections, symmetric instability, potential vorticity, cyclone intensity or computation of entropy production\ldots

It should be noted that, while it is possible and no doubt interesting to study the equation of entropy and $\theta_s$ with application of the third principle of thermodynamics, this does not call into question the Navier-Stokes equations.
However, it would be better to take advantage of the smoother appearance of $\theta_s$ in space, with lower gradients, which could reduce numerical errors for advection patterns. 
It would also be a question of being able to study, by reinforcing or invalidating them, the principles of ``maximum entropy production'' or ``maximum entropy'', two principles that some studies suspect of controlling the behaviour of the atmosphere and climate systems.

More concrete research actions have been launched in recent years and seem promising. 
Turbulent flows over ocean surfaces appear to have a much better fit with entropy and $\theta_s$ than with $\theta_l$ (Marquet and Belamari, 2017).
The exchange coefficients, which are the ``stiffness'' of the restoring forces with which the turbulence acts on the different variables, seem to differ near the surface and in altitude (Marquet et al., 2017).
In keeping with Richardson's (1922, p.177) remark, this would call into question a hypothesis used in all current atmospheric turbulence schemes.
Indeed, this could explain why well-mixed zones in entropy and  $\theta_s$ may exist whereas gradients of $q_t$ persist, and this may be explained if the Lewis number is larger than the unit, i.e. if the exchange coefficient for entropy is greater than that for humidity.

 \section{\underline{\Large References}}
\label{section Biblio}
\vspace{-1mm}

\noindent
$\bullet$ Bauer, L.A., (1908). 
The relation between ``potential temperature'' and ``entropy''. 
{\it Phys. Rev.\/}, 26, Series I, 177--183. 
(see~: 
``The Mechanics of the Earth Atmosphere, 
a collection of translations by Cleveland Abbe'', (1910). 
Smithsonian Miscellaneous Collections. 
Art. XXII, 495--500).

\noindent
$\bullet$ Betts A.K., (1973).  
Non-precipitating cumulus convection and its parameterization. 
{\it Quart. J. Roy. Meteorol. Soc.\/}, 99, 178--196.

\noindent
$\bullet$ von Bezold, W., (1888a). 
Zur Thermodynamik der Atmosphere. 
{\it Sitzungsberichte der Königl. Preuss. Akademie der Wissenschaften zu Berlin\/}, 
21, 485--52 
(see~: {\it ``On the thermo-dynamics of the atmosphere (first communucation)''\/} 
in ``The Mechanics of the Earth Atmosphere, 
a collection of translations by Cleveland Abbe'', (1891). 
Smithsonian Miscellaneous Collections. Art. XV, 212--242).

\noindent
$\bullet$ von Bezold, W., (1888b). 
Zur Thermodynamik der Atmosphere. 
{\it Sitzungsberichte der Königl. Preuss. Akademie der Wissenschaften zu Berlin\/}, 
46, 1189--1206 
(see~: {\it ``On the thermo-dynamics of the atmosphere (second communucation)\/}'' 
in ``The Mechanics of the Earth Atmosphere, 
a collection of translations by Cleveland Abbe'', (1891). 
Smithsonian Miscellaneous Collections. Art. XVI, 243--256).

\noindent
$\bullet$ Bjerknes, V., (1904), 
Das Problem der Wettervorhersage, betrachtet vom Standpunkte der Mechanik und der Physik. 
{\it Meteor. Zeit.\/}, 21, 1--7 
(see the English translation by E. Volken et S. Brönnimann, (2009)\,: 
``The problem of weather forecast, considered from the point of view of mechanics and physics''
{\it Meteor. Zeit.\/}, 18, 663--667).

\noindent
$\bullet$ 
Bretherton, C., Pincus, R., (1995). Cloudiness and marine boundary layer dynamics in the ASTEX lagrangian experiment. Part I : synoptic setting and vertical structure. {\it J. Atmos. Sci.\/}, 52, 2707--2723.

\noindent
$\bullet$ 
Dauhut, T., Chaboureau, J.-P. , Mascart ,P., Pauluis O., (2017). The atmospheric overturning induced by Hector the Convector. {\it J. Atmos. Sci.\/}, 74, 3271--3284.

\noindent
$\bullet$ 
Deardorff, J.W., (1980). Cloud top entrainment instability. {\it J. Atmos. Sci.\/}, 37, 131--147

\noindent
$\bullet$ Emanuel, K.A., (1994). 
{\it Atmospheric convection\/}. 
Oxford University Press. 580~p.

\noindent
$\bullet$ Ertel, H., (1942a). 
Ein neuer hydrodynamischer Wirbelsatz (A new hydrodynamic vorticity theorem). 
{\it Meteorol. Zeit.\/}, 59, 277--281.

\noindent
$\bullet$ Ertel, H., (1942b). 
Über hydrodynamische Wirbelsätze (About hydrodynamic vortex theorems). 
{\it Physik. Zeit. Leipzig\/}, 43, 526--529.

\noindent
$\bullet$ Hertz, H., (1884). 
Graphische Methode zur Bestimmung der adiabatischen Zustsänderungen feuchter Luft.
{\it Meteor. Zeit.\/}, 1, 421--431 ; 320--338 
(see the English translation~: 
{\it ``A graphic method of determining the adiabatic changes in the condition of moist air''\/}
in ``The Mechanics of the Earth Atmosphere, 
a collection of translations by Cleveland Abbe'', (1891). 
Smithsonian Miscellaneous Collections. Art.XIV, 198--211).

\noindent
$\bullet$ Hauf, T., Höller, H., (1987). 
Entropy and potential temperature.
{\it J. Atmos. Sci.\/}, 44, 2887--2901.

\noindent
$\bullet$ von Helmholtz, H., (1888). 
Über atmosphaerische Bewegungen. 
{\it Sitzungsberichte der König. Preuss. Akademie der Wissenschaften zu Berlin\/}, 
26, 647--663 
(see the English translation~: 
{\it ``On atmospheric movements''\/} in 
``The Mechanics of the Earth Atmosphere,
 a collection of translations by Cleveland Abbe'', (1891). 
 Smithsonian Miscellaneous Collections. Art.V, 78--93).

\noindent
$\bullet$ Joules. J.P., (1845). 
On the changes of temperature produced by the rarefaction and condensation of air. 
{\it Phil. Mag.\/}, Series 3, Vol 26, number 74, 369--383.

\noindent
$\bullet$ Knoche, W.,  (1906), 
{\it Ueber die räumliche und zeitliche Verteilung des Wärmegehalts der unteren Luftschicht
(On the spatial and temporal distribution of the heat content of the lower  layer or air)\/}. 
PhD-thesis. Friedrich-Wilhelms-Universität. 46~p.

\noindent
$\bullet$ 
MacVean, M.K., Mason, P.J., (1990). Cloud-top entrainment instability through small-scale mixing and its parameterization in numerical models. {\it J. Atmos. Sci.\/}, 47, 1012--1030.

\noindent
$\bullet$ Marquet, P., (2011). 
Definition of a moist entropy potential temperature: application to FIRE-I data flights.
{\it Quart. J. Roy. Meteorol. Soc.\/}, 137, 768--791. 
\url{http://arxiv.org/abs/1401.1097}

\noindent
$\bullet$ Marquet, P., (2014). 
On the definition of a moist-air potential vorticity. 
{\it Quart. J. Roy. Meteorol. Soc.\/}, 140, 917--929. 
\url{http://arxiv.org/abs/1401.2006}

\noindent
$\bullet$ Marquet, P., (2016). 
{\it Étude de l'énergétique de l'air humide et des paramétrisations physiques de l'atmosphère : propriétés de l'exergie, de l'enthalpie utilisable, de l'entropie et de l'enthalpie
(Study of the energy of moist air and physical parameterizations of the atmosphere: properties of exergy, available enthalpy, entropy and enthalpy)\/}.
Habilitation memoir of the Institut National Polytechnique (INP) of Toulouse, 311~p. 
\url{https://tel.archives-ouvertes.fr/tel-01504276}

\noindent
$\bullet$ Marquet, P., (2017). 
A Third-Law Isentropic Analysis of a Simulated Hurricane. 
{\it J. Atmos. Sci.\/}, 74, 3451--3471. 
\url{https://arxiv.org/abs/1704.06098}

\noindent
$\bullet$ Marquet, P., (2019). 
Le troisième principe ou une definition absolue de l'entropie. 
Partie~1~: les origines et applications en thermodynamique. 
{\it La M\'et\'eorologie\/}, Accepted for publication.
\url{http://documents.irevues.inist.fr/handle/2042/14834}
The third law of thermodynamics or an absolute definition for Entropy. 
Part~1: the origin and applications in thermodynamics.
\url{https://arxiv.org/abs/1904.11696}

\noindent
$\bullet$ 
Marquet, P., Belamari, S., (2017). On new bulk formulas based on moist-air entropy. {\it CAS/JSC, WGNE, WCRP Report\/}. http://bluebook.meteoinfo

\noindent
$\bullet$ Marquet, P., Dauhut, Th., (2018). 
Reply to ``Comments on A Third-Law Isentropic Analysis of a Simulated Hurricane''. 
{\it J. Atmos. Sci.\/}, 75, 3735--3747. 
\url{https://arxiv.org/abs/1805.00834}

\noindent
$\bullet$ Marquet, P., Geleyn J.-F., (2013). 
On a general definition of the squared Brunt-Väisälä frequency associated with the specific moist entropy potential temperature. 
{\it Quart. J. Roy. Meteorol. Soc.\/}, 139, 85--100. 
\url{http://arxiv.org/abs/1401.2379}

\noindent
$\bullet$ Marquet P., Geleyn J.-F., (2015). 
{\it Formulations of moist thermodynamics for atmospheric modelling\/}. 
Parameterization of Atmospheric Convection. 
Vol~II: Current Issues and New Theories. 
Plant and Yano Ed., Imperial College Press, 221--274. 
\url{http://arxiv.org/abs/1510.03239}

\noindent
$\bullet$ Mrowiec, A.,  Pauluis, O.,  Zhang, F.,  (2016).  
Isentropic Analysis of a Simulated Hurricane. 
{\it J. Atmos. Sci.\/}, 73, 1857--1870.

\noindent
$\bullet$ Marquet, P., Maurel, W., Honnert R., (2017). 
On consequences of measurements of turbulent Lewis number 
from observations. 
CAS/JSC, WGNE, WCRP Report.
\url{http://bluebook.meteoinfo}

\noindent
$\bullet$ Normand, C.W.B., (1921). 
Wet Bulb Temperatures and the Thermodynamics of the Air.  
{\it Memoirs of the India Meteorological Department.  
Meteorological Office\/}, 23, 1--22.

\noindent
$\bullet$ Pauluis, O., (2011). 
Water Vapor and Mechanical Work: A Comparison of Carnot and Steam Cycles, 
{\it J. Atmos. Sci.\/}, 68, 91--102.

\noindent
$\bullet$ Pauluis, O., Czaja, A., Korty, R.,  (2010).  
The global atmospheric circulation in moist isentropic coordinates. 
{\it J. Climate\/}, 23,  3077--3093.

\noindent
$\bullet$ Poisson, S.D., (1833). 
{\it Traité de mécanique (Treatise of Mechanics)\/}. 
Second Edition,  
Bachelier, Paris,
782~p.

\noindent
$\bullet$
Randall, D.A., (1980). Conditional instability of the first kind upside-down. {\it J. Atmos. Sci.\/}, 37, 125--130.

\noindent
$\bullet$ Richardson, L.F.,  (1919).  
Atmospheric stirring measured by precipitation. 
{\it Proc. Roy. Soc. London (A)\/}.  96,  9--18.

\noindent
$\bullet$ Richardson, L.F.,  (1920).  
The Supply of Energy from and to Atmospheric Eddies. 
{\it Proc. Roy. Soc. London (A)\/}.  97,  354--373.

\noindent
$\bullet$ Richardson, L.F.,  (1922).  
{\it Weather prediction by numerical process\/}. 
Cambridge University Press. 229~p.

\noindent
$\bullet$ Rossby, C.-G., (1932).  
{\it Thermodynamics applied to air mass analysis\/}. 
Papers in Physical Oceanography and Meteorology, 
Vol 1, number 3. 
Cambridge, Massachusetts, 57~p.

\noindent
$\bullet$ Saunders, P.M., (1957). 
The thermodynamics of saturated air : 
a contribution to the classical theory. 
{\it Quart. J. Roy. Meteorol. Soc.\/}, 83, 342--350.

\noindent
$\bullet$ Schubert, J., (1904). 
{\it Der Wärmeaustausch im festen Erdboden, in Gewässern und in der Atmosphäre 
(The heat exchange in solid earth, waters and the atmosphere)\/}. 
Verlag Springler, Berlin. 30~p.

\noindent
$\bullet$ Schubert et al., (2004). 
English translations of twenty-one of Ertel's papers on geophysical fluid dynamics.
{\it Meteorol. Zeit.\/}, 13 , 527--576.

\noindent
$\bullet$ Stevens, B., Siebesma, P., (2019). 
{\it Clouds as Fluids\/}. 
In: ``Clouds and Climate: Climate Sciences Greatest Challenge''. 
Siebesma, Bony, Jakob and Stevens Eds, 
Cambridge Univ. Press., Cambridge UK (in press).

\noindent
$\bullet$ Thomson, W., (1862). 
On the convective equilibrium of temperature in the atmosphere. 
{\it Manch. Lit. Philos. Soc.\/}, 2, 170--176.

  \end{document}